\documentclass[11pt]{article}

\usepackage{hyperref,amsfonts,amsmath,amssymb,bm,a4wide,cite,subfigure,graphicx}

\begin{document}

\title{\textbf{Impact of antiparticle degrees of freedom on neutrino flavor oscillations
in frames of quantum field theory}}

\author{Maxim Dvornikov\thanks{maxim.dvornikov@gmail.com}
\\
\small{\ Pushkov Institute of Terrestrial Magnetism, Ionosphere} \\
\small{and Radiowave Propagation (IZMIRAN),} \\
\small{108840 Moscow, Troitsk, Russia}}
%\\
%\small{\ Bogoliubov Laboratory of Theoretical Physics,} \\
%\small{Joint Institute for Nuclear Research,} \\
%\small{141980 Dubna, Moscow region, Russia}}

\date{}

\maketitle

\begin{abstract}
We study neutrino flavor oscillations using the approach based on
the quantum field theory (QFT), where neutrinos are taken to be virtual
particles. One deals with the propagators of neutrino mass eigenstates
in this formalism. Previously, while applying this approach to neutrino
oscillations in external fields, we decomposed the propagators and
used only the particle contribution in the calculation of the matrix
element. In the present work, we carefully justify the validity of
this kind of transformation by considering neutrino oscillations in
vacuum. In principle, the results obtained can be extended for the
QFT applied to neutrino oscillations in external fields.
\end{abstract}

In the wake of numerous experiments (see, e.g., Ref.~\cite{Fuk98,Ahm02}),
neutrinos were established to be massive particles having nonzero
mixing between different flavors. These neutrino properties result
in transitions between neutrino types which are called neutrino flavor
oscillations. Historically, neutrino oscillations are described using
the quantum mechanical approach (see, e.g., Ref.~\cite{Bil15}).
However, this kind of formalism was shown in Ref.~\cite{NauNau20}
to have certain shortcomings.

To avoid some of the shortcomings, the approach, where neutrino states are represented as noncovariant wave packets, was proposed (see, e.g., Ref.~\cite{GiuKim07}). The covariant generalization of the neutrino wave packets formalism was developed in Ref.~\cite{NauNau20}. Nevertheless, 
an approach to scrutinize neutrino oscillations, consistent with the quantum fields theory (QFT), has to be developed. This kind of formalisms was developed in Refs.~\cite{Kob82,GiuKimLee93,GriSto96}. The originally proposed QFT approach was significantly extended in Ref.~\cite{NauNau20}, e.g.,  by the consideration of the covariant wave packets of in- and out-states of particles involved in a reaction.

\begin{figure}
  \centering
  \includegraphics[scale=1]{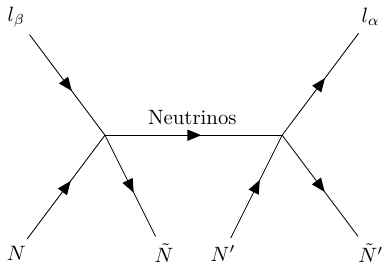}
  \protect
  \caption{The schematic illustration of the process which corresponds to neutrino flavor oscillations in frames of QFT.}\label{fig:feyndiag}
\end{figure}

As a rule, the QFT approach incorporates both the production, the propagation, and the detection
of neutrinos. Neutrinos are supposed in Ref.~\cite{Kob82} to be produced in a source, which
is a heavy nucleus $N$, in a reaction with the incoming charged lepton
$l_{\beta}$, $l_{\beta}+N\to\tilde{N}+\text{neutrinos}$. Then, these
neutrinos propagate in space and interact with a detector, which again
is a heavy nucleus $N'$. A charged lepton $l_{\alpha}$ is produced
in the wake of this interaction, $\text{neutrinos}+N'\to\tilde{N}'+l_{\alpha}$.
If $l_{\alpha}\neq l_{\beta}$, the process in question is interpreted
as neutrino oscillations. Note that neutrinos are taken to be virtual
particles in this approach. The described process is schematically
depicted in Fig.~\ref{fig:feyndiag}.

In Refs.~\cite{Dvo25,Dvo25b}, we extended the above picture of neutrino
oscillations to include various external fields which neutrinos can
interact with. In particular, we considered the cases of the neutrino
interaction with background matter and with an external magnetic field.
Note that neutrino oscillations in external fields in frames of QFT
were also studied in Refs.~\cite{CarChu99,AkhWil13,EgoVol22}.

The main purpose of the present work is to justify one assumption
made in Refs.~\cite{Dvo25,Dvo25b}. As we shall see shortly, a process
interpreted as neutrino oscillations involves the propagators of virtual
neutrinos. In Refs.~\cite{Dvo25,Dvo25b}, these propagators were
decomposed into particle and antiparticle parts, with only one them
being used in the computation of the matrix element. The remaining
part was claimed to give a negligible contribution to the matrix element
for ultrarelativistic neutrinos. It is an expected but not so obvious
assumption. Now, we demonstrate the validity of this statement considering
neutrino oscillations in vacuum.

In frames of QFT, the following $S$-matrix element corresponds to
neutrino flavor oscillations,%\foreignlanguage{english}{
\begin{align}\label{eq:Smatr}
  S = & -\frac{1}{2}\left(\sqrt{2}G_{\mathrm{int}}\right)^{2}
  \int\mathrm{d}^{4}x\mathrm{d}^{4}y
  \notag
  \\
  & \times
  \left\langle
    \tilde{N},\tilde{N}',l_{\alpha}
    \left|
      T
      \left\{
        j_{\mu}^{\dagger}(x)J^{\mu}(x)j^{\nu}(y)J_{\nu}^{\dagger}(y)
      \right\}
    \right|N,N',l_{\beta}
  \right\rangle,
\end{align}
where
\begin{equation}
j_{\mu}=\sum_{\lambda}\bar{\nu}_{\lambda\mathrm{L}}\gamma_{\mu}l_{\lambda\mathrm{L}},\label{eq:lepcurr}
\end{equation}
is the operator valued leptonic current, $J^{\mu}$ is the nuclear
current operator, and $G_{\mathrm{int}}$ is the coupling constant,
and $\gamma^{\mu}=(\gamma^{0},\bm{\gamma})$ are the Dirac matrices.
In Eq.~(\ref{eq:lepcurr}), the operators of flavor neutrinos, $\nu_{\lambda}$,
and of charged leptons, $l_{\lambda}$, stay as the left chiral projections
$\propto\tfrac{1}{2}(1-\gamma^{5})$, where $\gamma^{5}=\mathrm{i}\gamma^{0}\gamma^{1}\gamma^{2}\gamma^{3}$.

The flavor neutrinos, which interact with other particles in the standard
model, do not have definite masses. To diagonalize the mass term in
the neutrino Lagrangian, one introduces the neutrino mass eigenstates
$\psi_{a}$, with the masses $m_{a}$, which are the superposition
of the neutrino flavor eigenstates. If we deal with only two neutrinos
for simplicity, e.g., $\nu_{e}$ and $\nu_{\mu}$, this matrix transformation
reads
\begin{equation}
\nu_{\lambda}=\sum_{a=1,2}U_{\lambda a}\psi_{a},\quad(U_{\lambda a})=\left(\begin{array}{cc}
\cos\theta & \sin\theta\\
-\sin\theta & \cos\theta
\end{array}\right),\label{eq:flmassrel}
\end{equation}
where $\theta$ is the vacuum mixing angle. 

Assuming that the nuclei in a source and a detector of neutrinos are
quite heavy, as well as that charged leptons propagate as plane waves,
one rewrites Eq.~(\ref{eq:Smatr}) as
\begin{equation}
  S=-2\pi G_{\mathrm{int}}^{2}\delta(E_{\alpha}-E_{\beta})
  e^{-i\mathbf{p}_{\alpha}\mathbf{x}_{2}+i\mathbf{p}_{\beta}\mathbf{x}_{1}}
  \mathrm{i}\mathcal{M}_{\beta\to\alpha},
\end{equation}
where
\begin{equation}
\mathcal{M}_{\beta\to\alpha}=\bar{u}_{\alpha}\gamma_{0}^{\mathrm{L}}\sum_{a}U_{\alpha a}U_{\beta a}^{*}\left(\int\frac{\mathrm{d}^{3}q}{(2\pi)^{3}}e^{\mathrm{i}\mathbf{qL}}S_{a}(E,\mathbf{q})\right)\gamma_{0}^{\mathrm{L}}u_{\beta}(p_{\beta}),\label{eq:matrel}
\end{equation}
is the matrix element, $u_{\beta}=u_{\beta}(p_{\beta})$ and $u_{\alpha}=u_{\alpha}(p_{\alpha})$
are the spinors of incoming and outgoing charged leptons having the
momenta $p_{\beta}^{\mu}=(E_{\beta},\mathbf{p}_{\beta})$ and $p_{\alpha}^{\mu}=(E_{\alpha},\mathbf{p}_{\alpha})$,
$E=(E_{\alpha}+E_{\beta})/2$, $\mathbf{L}$ is the vector connecting
the positions of a source $\mathbf{x}_{1}$ and a detector $\mathbf{x}_{2}$,
and
\begin{equation}
S_{a}(q^{\mu})=\frac{\gamma_{\mu}q^{\mu}+m_{a}}{q^{2}-m_{a}^{2}+\mathrm{i}0},\label{eq:vacpropgen}
\end{equation}
is the 4D Fourier image of the vacuum propagator of a neutrino mass
eigenstate $S_{a}(x-y)=-\mathrm{i}\left\langle 0\left|T\left\{ \psi_{a}(x)\bar{\psi}_{a}(y)\right\} \right|0\right\rangle $.

We cast the propagator to the form, $S_{a}=S_{a}^{(\psi)}+S_{a}^{(\bar{\psi})}$,
where
\begin{align}
S_{a}^{(\psi)} & =\frac{1}{2(q_{0}-E_{a}+\mathrm{i}0)}\left(\gamma^{0}-\frac{1}{E_{a}}\bm{\gamma}\mathbf{q}+\frac{m_{a}}{E_{a}}\right),\label{eq:propnu}\\
S_{a}^{(\bar{\psi})} & =\frac{1}{2(q_{0}+E_{a}-\mathrm{i}0)}\left(\gamma^{0}+\frac{1}{E_{a}}\bm{\gamma}\mathbf{q}-\frac{m_{a}}{E_{a}}\right).\label{eq:propantinu}
\end{align}
Here $E_{a}=\sqrt{q^{2}+m_{a}^{2}}$ is the energy of a massive neutrino.
The propagator $S_{a}^{(\psi)}$ in Eq.~(\ref{eq:propnu}) has the
positive imaginary term $+\mathrm{i}0$ in the denominator. We referred
to such a function in Ref.~\cite{Dvo25} as to the propagator of
particles. On the contrary, $S_{a}^{(\bar{\psi})}$ in Eq.~(\ref{eq:propantinu})
has the negative imaginary term $-\mathrm{i}0$ in the denominator.
We can attribute it to the antiparticle propagator. 

We calculate the contributions of both neutrinos and antineutrinos
to the matrix element in Eq.~(\ref{eq:matrel}) by replacing $S_{a}\to S_{a}^{(\psi)}$
and $S_{a}\to S_{a}^{(\bar{\psi})}$. To simplify the calculations
we adopt the forward scattering approximation for leptons by taking
that $u_{\alpha,\beta}^{\mathrm{L}}=(0,0,0,1)^{\mathrm{T}}$. It means that both
leptons are left polarized and propagate along the positive direction
of the $z$-axis. We rely on the Dirac matrices in the chiral representation
here. We also suppose that $\mathbf{L}=L\mathbf{e}_{z}$.

First, we study the particles contributions to the matrix element.
Using Eq.~(\ref{eq:propnu}), one gets that
\begin{equation}
\bar{u}_{\alpha}\gamma_{0}^{\mathrm{L}}\left(\gamma^{0}-\frac{1}{E_{a}}\bm{\gamma}\mathbf{q}+\frac{m_{a}}{E_{a}}\right)\gamma_{0}^{\mathrm{L}}u_{\beta}(p_{\beta})=\frac{1}{2}\left(1+\frac{q_{z}}{E_{a}}\right).\label{eq:numav}
\end{equation}
Then, we use the cylindrical coordinates for the momentum of the virtual
neutrino, $\mathbf{q}=\rho\mathbf{e}_{\rho}+z\mathbf{e}_{z}$. Taking
into account Eq.~(\ref{eq:numav}), the integral in Eq.~(\ref{eq:matrel}),
averaged over leptonic states, becomes
\begin{align}\label{eq:I1}
  I_{\psi} = & \bar{u}_{\alpha}\gamma_{0}^{\mathrm{L}}
  \left(
    \int\frac{\mathrm{d}^{3}q}{(2\pi)^{3}}e^{\mathrm{i}\mathbf{qL}}S_{a}^{(\psi)}(E,\mathbf{q})
  \right)
  \gamma_{0}^{\mathrm{L}}u_{\beta}(p_{\beta})
  \notag
  \\
  & =
  \frac{1}{8\pi^{2}}
  \int_{0}^{\infty}\rho\mathrm{d}\rho
  \int_{-\infty}^{+\infty}\mathrm{d}z
  \frac{
  \left(
    1+\frac{z}{E_{a}}
  \right)
  e^{\mathrm{i}zL}}{E-E_{a}+\mathrm{i}0}.
\end{align}
The integral over $z$ in Eq.~(\ref{eq:I1}) is computed in the complex
plane. Assuming that $L>0$, we should close the contour in the upper
half plane to provide the integral convergence. The poles of the integrand,
which are the solution of the equation $E-E_{a}+\mathrm{i}0=0$, have
the following form:
\begin{align}
z_{0} & =\begin{cases}
\sqrt{\rho_{0}^{2}-\rho^{2}}+\mathrm{i}0, & \text{if}~\rho<\rho_{0},\\
\mathrm{i}\sqrt{\rho^{2}-\rho_{0}^{2}}, & \text{if}~\rho>\rho_{0},
\end{cases}\label{eq:polesnu}
\end{align}
where $\rho_{0}=\sqrt{E-m_{a}^{2}}$.

Decomposing the denominator of the integrand in Eq.~(\ref{eq:I1}) near
the pole, 
\begin{equation}
E-E_{a}+\mathrm{i}0=-(z-z_{0})\left.\frac{\mathrm{d}E_{a}}{\mathrm{d}z}\right|_{z=z_{0}}+\dotsb=-(z-z_{0})\frac{z_{0}}{E},\label{eq:decompnu}
\end{equation}
since $E_{a}(z_{0})=E$, one obtains that
\begin{align}\label{eq:I2}
  I_{\psi} = & -\frac{\mathrm{i}}{4\pi}
  \bigg[
    \int_{0}^{\rho_{0}}\rho\mathrm{d}\rho e^{\mathrm{i}L\sqrt{\rho_{0}^{2}-\rho^{2}}}
    \left(
      1+\frac{E}{\sqrt{\rho_{0}^{2}-\rho^{2}}}
    \right)
    \notag
    \\
    & +
    \int_{\rho_{0}}^{\infty}\rho\mathrm{d}\rho e^{-L\sqrt{\rho^{2}-\rho_{0}^{2}}}
    \left(
      1-\frac{\mathrm{i}E}{\sqrt{\rho^{2}-\rho_{0}^{2}}}
    \right)
  \bigg].
\end{align}
The remaining integrals over $\rho$ in Eq.~(\ref{eq:I2}) are computed
with help of the expressions,
\begin{align}
  \int_{0}^{\rho_{0}}\rho\mathrm{d}\rho\frac{e^{\mathrm{i}L\sqrt{\rho_{0}^{2}-\rho^{2}}}}{\sqrt{\rho_{0}^{2}-\rho^{2}}} & =
  -\frac{\mathrm{i}}{L}(e^{\mathrm{i}\rho_{0}L}-1),
  \label{eq:basint1}
  \\
  \int_{0}^{\rho_{0}}\rho\mathrm{d}\rho e^{\mathrm{i}L\sqrt{\rho_{0}^{2}-\rho^{2}}} & =
  -\frac{\mathrm{i}}{L^{2}}e^{\mathrm{i}\rho_{0}L}(L\rho_{0}+\mathrm{i})-\frac{1}{L^{2}},
  \label{eq:basint2}
  \\
  \int_{\rho_{0}}^{\infty}\rho\mathrm{d}\rho\frac{e^{-L\sqrt{\rho^{2}-\rho_{0}^{2}}}}{\sqrt{\rho^{2}-\rho_{0}^{2}}} & =
  \frac{1}{L},
  \\
  \int_{\rho_{0}}^{\infty}\rho\mathrm{d}\rho e^{-L\sqrt{\rho^{2}-\rho_{0}^{2}}} & =
  \frac{1}{L^{2}}.
  \label{eq:basint}
\end{align}
Eventually, one gets that
\begin{equation}
I_{\psi}=-\frac{Ee^{\mathrm{i}\sqrt{E-m_{a}^{2}}L}}{4\pi L}\left(1+\frac{\sqrt{E-m_{a}^{2}}}{E}+\frac{\mathrm{i}}{LE}\right).\label{eq:I3}
\end{equation}
As a rule, one considers the situation when the propagation distance
is great, $L\gg E^{-1}$. Thus, we can neglect the last term in Eq.~(\ref{eq:I3}).

Using Eqs.~(\ref{eq:flmassrel}) and~(\ref{eq:I3}) and considering
the two neutrinos system, one gets the particle contribution to the
matrix element in the form,
\begin{align}
\mathcal{M}_{e\to\mu}^{(\psi)}\approx & -\frac{Ee^{\mathrm{i}EL}}{4\pi L}\sin2\theta\left[e^{-\mathrm{i}\frac{m_{2}^{2}L}{2E}}\left(1-\frac{m_{2}^{2}}{4E^{2}}\right)-e^{-\mathrm{i}\frac{m_{1}^{2}L}{2E}}\left(1-\frac{m_{1}^{2}}{4E^{2}}\right)\right].\nonumber \\
 & =-\frac{Ee^{\mathrm{i}EL}}{4\pi L}\sin2\theta\left[e^{-\mathrm{i}\frac{m_{2}^{2}L}{2E}}-e^{-\mathrm{i}\frac{m_{1}^{2}L}{2E}}+\mathcal{O}\left(\frac{m_{a}^{2}}{E^{2}}\right)\right],\label{eq:martelnu}
\end{align}
where we assume that neutrinos are ultrarelativistic with $m_{a}\ll E$.

Now, we compute the antiparticle contribution to the matrix element.
For this purpose, we substitute $S_{a}^{(\bar{\psi})}$ in Eq.~(\ref{eq:propantinu})
to Eq.~(\ref{eq:matrel}). We just list the main modifications in
Eqs.~(\ref{eq:numav})-(\ref{eq:I3}). Equation~(\ref{eq:numav})
becomes
\begin{equation}
\bar{u}_{\alpha}\gamma_{0}^{\mathrm{L}}\left(\gamma^{0}+\frac{1}{E_{a}}\bm{\gamma}\mathbf{q}-\frac{m_{a}}{E_{a}}\right)\gamma_{0}^{\mathrm{L}}u_{\beta}(p_{\beta})=\frac{1}{2}\left(1-\frac{q_{z}}{E_{a}}\right).\label{eq:antinumav}
\end{equation}
The integral in the matrix element reads
\begin{align}\label{eq:I1anti}
  I_{\bar{\psi}} = & \bar{u}_{\alpha}\gamma_{0}^{\mathrm{L}}
  \left(
    \int\frac{\mathrm{d}^{3}q}{(2\pi)^{3}}e^{\mathrm{i}\mathbf{qL}}S_{a}^{(\bar{\psi})}(E,\mathbf{q})
  \right)
  \gamma_{0}^{\mathrm{L}}u_{\beta}(p_{\beta}) 
  \notag
  \\
  & =
  \frac{1}{8\pi^{2}}
  \int_{0}^{\infty}\rho\mathrm{d}\rho
  \int_{-\infty}^{+\infty}\mathrm{d}z\frac{
  \left(
    1-\frac{z}{E_{a}}
  \right)
  e^{\mathrm{i}zL}}{E+E_{a}-\mathrm{i}0}.
\end{align}
Instead of Eq.~(\ref{eq:polesnu}), the poles of the integrand in
Eq.~(\ref{eq:I1anti}) are
\begin{align}
\bar{z}_{0} & =\begin{cases}
-\sqrt{\rho_{0}^{2}-\rho^{2}}+\mathrm{i}0, & \text{if}~\rho<\rho_{0},\\
\mathrm{i}\sqrt{\rho^{2}-\rho_{0}^{2}}, & \text{if}~\rho>\rho_{0}.
\end{cases}\label{eq:polesantinu}
\end{align}
Note that we close the contour also in the upper half plane while
integrating over $z$. The decomposition in Eq.~(\ref{eq:decompnu})
takes the form,
\begin{equation}
E+E_{a}-\mathrm{i}0=(z-\bar{z}_{0})\left.\frac{\mathrm{d}E_{a}}{\mathrm{d}z}\right|_{z=\bar{z}_{0}}+\dotsb=-(z-\bar{z}_{0})\frac{\bar{z}_{0}}{E},\label{eq:decompantinu}
\end{equation}
since $E_{a}(\bar{z}_{0})=-E$.

The integral in Eq.~(\ref{eq:I2}) is rewritten as
\begin{align}\label{eq:I2antinu}
  I_{\bar{\psi}} = & \frac{\mathrm{i}}{4\pi}
  \bigg[
    \int_{0}^{\rho_{0}}\rho\mathrm{d}\rho e^{-\mathrm{i}L\sqrt{\rho_{0}^{2}-\rho^{2}}}
    \left(
      1-\frac{E}{\sqrt{\rho_{0}^{2}-\rho^{2}}}
    \right)
    \notag
    \\
    & +
    \int_{\rho_{0}}^{\infty}\rho\mathrm{d}\rho e^{-L\sqrt{\rho^{2}-\rho_{0}^{2}}}
    \left(
      1-\frac{\mathrm{i}E}{\sqrt{\rho^{2}-\rho_{0}^{2}}}
    \right)
  \bigg].
\end{align}
The integrations in Eq.~(\ref{eq:I2antinu}) can be carried out if
we replace $L\to-L$ in Eqs.~(\ref{eq:basint1}) and~(\ref{eq:basint2}).
The final result for $I_{\bar{\psi}}$ reads
\begin{equation}
I_{\bar{\psi}}=\frac{Ee^{-\mathrm{i}\sqrt{E-m_{a}^{2}}L}}{4\pi L}\left(1-\frac{\sqrt{E-m_{a}^{2}}}{E}+\frac{\mathrm{i}}{LE}\right).\label{eq:I3antinu}
\end{equation}
Using Eq.~(\ref{eq:I3antinu}), one gets the antiparticle contribution
to the matrix element,
\begin{align}
\mathcal{M}_{e\to\mu}^{(\bar{\psi})}\approx & \frac{Ee^{-\mathrm{i}EL}}{4\pi L}\sin2\theta\left[\frac{m_{2}^{2}}{4E^{2}}e^{\mathrm{i}\frac{m_{2}^{2}L}{2E}}-\frac{m_{1}^{2}}{4E^{2}}e^{-\mathrm{i}\frac{m_{1}^{2}L}{2E}}\right],\label{eq:martelantinu}
\end{align}
where we are based on the same assumptions as for the derivation of
Eq.~(\ref{eq:martelnu}).

Comparing Eqs.~(\ref{eq:martelnu}) and~(\ref{eq:martelantinu}),
one obtains that
\begin{equation}
\frac{|\mathcal{M}_{e\to\mu}^{(\bar{\psi})}|}{|\mathcal{M}_{e\to\mu}^{(\psi)}|}\sim\frac{m_{a}^{2}}{E^{2}}\ll1.
\end{equation}
Thus, the antiparticle part of the propagator in Eq.~(\ref{eq:propantinu})
gives a contribution to the matrix element which is negligible for
ultrarelativistic neutrinos.

Finally, we demonstrate that the matrix element in Eq.~(\ref{eq:martelnu})
results in the appropriate transition probability expression. Based
on the fact that $P_{\nu_{e}\to\nu_{\mu}}\propto|\mathcal{M}_{e\to\mu}|^{2}$,
one gets that
\begin{equation}
P_{\nu_{e}\to\nu_{\mu}}(L)=\sin^{2}2\theta\sin^{2}\left(\frac{\Delta m^{2}}{4E}L\right),\label{eq:trprob}
\end{equation}
where $\Delta m^{2}=m_{2}^{2}-m_{1}^{2}$. One can see that Eq.~(\ref{eq:trprob})
reproduces the known expression for the transition probability of
neutrino oscillations in vacuum (see, e.g., Ref.~\cite{Bil15}).

In conclusion, we mention that we have analyzed the validity of the
decomposition of the propagators of neutrino mass eigenstates which
was used in Ref.~\cite{Dvo25}. After the separation of the particle
and antiparticle parts in the propagator, one uses only the particle
part in the formalism for the description of neutrino flavor oscillations
based on QFT. We have shown that the contribution of the antiparticle
part to the matrix element is suppressed for ultrarelativistic neutrinos.
This fact has been demonstrated for neutrino oscillations in vacuum
since one can carry out all the calculations analytically in this
case. Despite the QFT based description of neutrino oscillations in
external fields is more complicated technically, one can extrapolate
the obtained result to this kind of situations.

\section*{Funding}

The research is supported by the government assignment of IZMIRAN.

\section*{Conflict of interest}

The author declares that there is no conflict of interest.


\begin{thebibliography}{20}

\bibitem{Fuk98}
  \emph{Fukuda~Y. et al.} [Super-Kamiokande Collaboration]
  Evidence for Oscillation of Atmospheric Neutrinos~//
  Phys. Rev. Lett.
  ---
  1998. ---
  \newblock V.~81. ---
  \newblock P.~1562--1567. ---  
  \newblock hep-ex/9807003.
    
\bibitem{Ahm02}
  \emph{Ahmad~Q.R. et al.} [SNO Collaboration]
  Direct Evidence for Neutrino Flavor Transformation from Neutral-Current Interactions in the Sudbury Neutrino Observatory~//
  Phys. Rev. Lett.
  ---
  2002. ---
  \newblock V.~89. ---
  \newblock P.~011301. ---   
  \newblock nucl-ex/0204008.

%\bibitem{Abu25}
%  \emph{Abubakar~S. et al.} [The NOvA Collaboration]
%  Precision measurement of neutrino oscillation parameters with 10 years of data from the NOvA experiment~//
%  \newblock arXiv:2509.04361.

\bibitem{Bil15}
  \emph{Bilenky~S.M.}
  Neutrino in Standard Model and beyond~//
  Phys. Part. Nucl.
  ---
  2015. ---
  \newblock V.~46. ---
  \newblock P.~475--496. ---
  \newblock arXiv:1501.00232.

\bibitem{NauNau20}
  \emph{Naumov~D.V., Naumov~V.A.}
  Quantum Field Theory of Neutrino Oscillations~//
  Phys. Part. Nucl.
  ---
  2020. ---
  \newblock V.~51. ---
  \newblock P.~1--106.

\bibitem{GiuKim07}
  \emph{Giunti~C., Kim~C.W.}
  Fundamentals of Neutrino Physics and Astrophysics.
  Oxford: Oxford University Press.
  2007.
  \newblock P.~299--321.

\bibitem{Kob82}
  \emph{Kobzarev~I.Yu., Martem'yanov~B.V., Okun'~L.B., Shchepkin~M.G.}
  Sum rules for neutrino oscillations~//
  Sov. J. Nucl. Phys.
  ---
  1982. ---
  \newblock V.~35. ---
  \newblock P.~708--712.

\bibitem{GiuKimLee93}
  \emph{Giunti~C., Kim~C.W., Lee~J.A.}
  On the treatment of neutrino oscillations without resort to weak eigenstates~//
  Phys. Rev. D.
  ---
  1993. ---
  \newblock V.~48. ---
  \newblock P.~4310--4317. ---
  \newblock hep-ph/9305276.

\bibitem{GriSto96}
  \emph{GrimusW., Stockinger~P.}
  Real oscillations of virtual neutrinos~//
  Phys. Rev. D.
  ---
  1996. ---
  \newblock V.~54. ---
  \newblock P.~3414--3419. ---
  \newblock hep-ph/9603430.

\bibitem{Dvo25}
  \emph{Dvornikov~M.}
  Quantum field theory treatment of neutrino flavor oscillations in matter~//
  Phys. Rev. D.
  ---
  2025. ---
  \newblock V.~111. ---
  \newblock P.~056009. ---
  \newblock arXiv:2411.19120.

\bibitem{Dvo25b}
  \emph{Dvornikov~M.}
  Quantum field theory treatment of the neutrino spin-flavor precession in a magnetic field~//
  \newblock arxiv:2504.14726.

\bibitem{CarChu99}
  \emph{Cardall~C.Y., Chung~D.J.H.}
  MSW effect in quantum field theory~//
  Phys. Rev. D.
  ---
  1999. ---
  \newblock V.~60. ---
  \newblock P.~073012. ---
  \newblock hep-ph/9904291.

\bibitem{AkhWil13}
  \emph{Akhmedov~E.Kh., Wilhelm~A.}
  Quantum field theoretic approach to neutrino oscillations in matter~//
  J. High Energy Phys.
  ---
  2013. ---
  \newblock V.~01. ---
  \newblock P.~165. ---
  \newblock arXiv:1205.6231.

\bibitem{EgoVol22}
  \emph{Egorov~V., Volobuev~I.}
  Quantum field-theoretical description of neutrino oscillations in magnetic field~//
  J. Exp. Theor. Phys.
  ---
  2022. ---
  \newblock V.~135. ---
  \newblock P.~197--208. ---
  \newblock arxiv:2107.11570.

\end{thebibliography}
\end{document}